\documentclass[12pt]{article}
\usepackage{amsfonts}
\begin{document}
\title{\bf
Integrable structure of the low-energy string gravity equations
in $D=4$ space-times\\ with two commuting isometries}
\author{
G.~A.~Alekseev${}^1$\footnote{e-mail: G.A.Alekseev@mi.ras.ru}~~and
M.~V.~Yurova${}^2$\footnote{e-mail: yurova@depni.sinp.msu.ru} 
\\
{\normalsize\emph{${}^1$ Steklov Mathematical Institute, Russian Academy of Sciences}}
\\[-0.5ex]
{\normalsize\emph{Gubkin St. 8, 119991, Moscow, Russia }}
\\[-0.5ex]
{\normalsize\emph{${}^2$ Institute of Nuclear Physics, Moscow State University}}
\\[-0.5ex]
{\normalsize\emph{Vorobjovy Gory, 119899 Moscow, Russia}}}
\date{}
\maketitle
\abstract{The generalized Einstein - Maxwell field equations which arise from a truncated bosonic part of the low - energy string gravity effective action in four dimensions (the so called Einstein-Maxwell - axion - dilaton theory) are considered. The integrable structure of these field equations for $D=4$ space-times with two commuting isometries is elucidated. We express the dynamical part of the reduced equations as integrability conditions of some overdetermined $4\times 4$-matrix linear system with a spectral parameter. The remaining
part of the field equations are expressed as the conditions of existence for this linear system of two $4\times 4$-matrix integrals of special structures.
This provides a convenient base for a generalization to these equations of various  solution generating methods developed earlier in General Relativity.}

\section{Introduction}
The generalized Einstein - Maxwell field equations considered below describes a  specifically coupled to gravity massless fields (axion, dilaton and one $U(1)$ gauge (``electromagnetic'') fields). The corresponding action (representing a truncated bosonic sectors of some low energy string gravity models in four dimensions) takes in Planck units the well known form:
\begin{equation}\label{Action}
{\cal S}=-\displaystyle\frac  1{16\pi}\int\{R{}^{(4)}-2\nabla_k\phi\nabla^k\phi
+e^{-2\phi} F_{ik} F^{ik}
-\displaystyle\frac 1{12} e^{-4\phi} H_{ijk} H^{ijk}\}\sqrt{-g} d^4\,x\nonumber
\end{equation}
where the indeces $i,j,k,\ldots$ are 4-dimensional, the metric $g_{ik}$ possess the ``most negative'' Lorentz signature, $\phi$ is the dilaton field and the Maxwell two-form $F_{ik}$ and a three-form $H_{ijk}$ are related to the gauge field $A_k$ and the antisymmetric tensor $B_{ik}$ by $F=d A$ and $H=d B-2 A\wedge F$.
In $D=4$ space-time, $F_{ik}$ and $H_{ijk}$ can be characterized also by their
dual fields ${\stackrel \ast F}_{ik}$ and $a^i$ respectively, and the field equations for the action (\ref{Action}) imply the existence of the ``magnetic'' vector $B_i$ and the scalar $a$ potentials for these dual fields ($\varepsilon^{ijkl}$ is the 4-dimensional Levi - Civita tensor):
$$
\begin{array}{lcl}
H_{ijk}=\varepsilon_{ijkl} a^l,&&F_{ik}=-\frac 12 \varepsilon_{ijkl}{\stackrel \ast F}{}^{kl},\\[1ex]
a_i=e^{4\phi}\nabla_i a,&&{\stackrel \ast F}_{ik}=e^{2\phi}\left[\partial_i B_k-\partial_k B_i+a (\partial_i A_k-\partial_k A_i)\right].
\end{array}
$$

Construction of exact solutions for the gravity model (\ref{Action}) and investigation of their physical and geometrical properties  has drawn much attention in the studies of the last decade. These studies revealed a nice internal symmetry structure of these field equations for space-times with two commuting isometries. In \cite{Bakas:1994} it was shown that in the particular case of vanishing  electromagnetic field these equations are integrable. In this case,  the dynamical variables and corresponding reduced equations split into two independent sectors (pure ``gravitational'' sector and the axion-dilatonic one) each governed by the Ernst $\sigma$-model equations. In the presence of $U(1)$ gauge field, the structure of the reduced dynamical equations can be expressed in the form of specific complex symmetric $2\times 2$-matrix generalization of the known vacuum Ernst equation in General Relativity. This type of $2\times 2$-matrix generalization of of the Ernst equation had been derived in \cite{Kumar-Ray:1995}. The symmetries of the dynamical equations for various effective string gravity models with isometries have been studied extensively (see, for example,  \cite{Schwarz-Sen}, \cite{Gal'tsov-Kechkin-Yurova} and the references therein). A number of attempts were made for the analysis of integrability and applications of some known solution generating methods for these equations. In particular, in \cite{Gal'tsov:1995} it was shown that a  Belinskii - Zakharov - like spectral problem can be associated with the reduced field equations corresponding to the action (\ref{Action}). Another kind of associated linear system has been suggested recently in \cite{Das-Maharana-Melikyan}. Some applications of the linear problem \cite{Gal'tsov:1995} for construction of particular form of soliton solutions have been considered in \cite{Yurova:2001}, \cite{Herrera-Aguilar-Mora-Luna}. Another way, based on the so called ``monodromy transform'' approach, was outlined in \cite{Alekseev:2003} (see also the references therein). One can notice, however, that in these studies the integrable structure of the equations under consideration has not been elaborated in all necessary details and any nontrivial solution generating methods have not been realized in these approaches.

For a construction of some effective solution generating methods one usually needs to reformulate the dynamical part of the field equations as 
some equivalent spectral problem, i.e. as integrability conditions of some associated linear system with spectral parameter, supplied with all necessary constraint conditions (such as, for example,  reality of some variables, given canonical structures, symmetric or coset structures of auxiliary matrix functions, etc.) also expressed more or less explicitly as the constraints on the solutions of this linear system. Such reformulation of the reduced dynamical equations for the action (\ref{Action}) for space-times with two commuting isometries, following the approach \cite{Alekseev:2003}, is the purpose of the present paper.

\section{Space-times with isometries}
We restrict our consideration of the gravity model (\ref{Action})
to the field configurations whose components and potentials all
are functions of two of the four space-time coordinates only.
These includes two physically important cases: the fields
depending on the time and only one spatial coordinate (the
hyperbolic case) or on two spatial coordinates only (the elliptic
case). Both of these cases are considered below in the same manner and
the only sign symbol $\epsilon$ ($\epsilon=1$ and $\epsilon=-1$ for the hyperbolic and the elliptic case respectively) will recall us about the
difference between them.
.

\subsection{Structure of the metric components}
The structure of the action (\ref{Action}) together with our
space-time symmetry conjecture imply that the space-time metric
can be considered in the block-diagonal form
\begin{equation}\label{Metric}
ds^2=g_{\mu\nu} dx^\mu dx^\nu+g_{ab} dx^a dx^b
\end{equation}
where $\mu,\nu=1,2$, $a,b,\ldots=3,4$ and the metric components
$g_{\mu\nu}$ and $g_{ab}$ depend on the coordinates $x^\mu$ and
not on $x^a$.\footnote{We do not specify here whether the
time-like coordinate is among the coordinates $x^\mu$, or it is
one of the ``ignorable'' coordinates $x^a$.} An appropriate choice
of the coordinates $x^\mu$ allows to present the metric components
$g_{\mu\nu}$ in a conformally flat form:
$$g_{\mu\nu}=f \eta_{\mu\nu},\qquad\eta_{\mu\nu}=
\left(\begin{array}{ll}\epsilon_1 & 0\\
0&\epsilon_2\end{array}\right),\qquad\begin{array}{l}\epsilon_1=\pm1,\\
\epsilon_2=\pm1\end{array}$$ where the conformal factor $f(x^\mu)
>0$. For $g_{ab}$ we use a parametrization
$$g_{ab}=\epsilon_0\left(\begin{array}{lcl}H&& H\Omega\\
H\Omega&&H\Omega^2+\displaystyle\frac {\epsilon\alpha^2}H\end{array}\right),
\qquad\epsilon_0=\pm1,\qquad \begin{array}{l}\det\Vert g_{ab}\Vert\equiv \epsilon\alpha^2,\\[1ex]
\epsilon\equiv -\epsilon_1\epsilon_2
\end{array}
$$
where the metric functions $\alpha(x^\mu)>0$, $H(x^\mu)>0$ and $\Omega(x^\mu)$ are introduced and the value of the sign symbol $\epsilon_0$ as well as the relation between the sign symbols $\epsilon_1$, $\epsilon_2$ and $\epsilon$ should provide the Lorentz signature of (\ref{Metric}).

\subsection{Geometrically defined coordinates}
It is easy to observe that in accordance with the field equations for the action (\ref{Action}), the function $\alpha$ should satisfy the
linear two-dimensional equation of the form $\eta^{\mu\nu}\partial_\mu\partial_\nu\alpha=0$ where the matrix $\eta^{\mu\nu}$ is inverse to $\eta_{\mu\nu}$. This is the d'Alambert  (for $\epsilon=1$) or Laplace (for $\epsilon=-1$) equation. It allows to introduce the function $\beta(x^\mu)$ ``harmonically'' conjugated  to
$\alpha$ using the relation: $\partial_\mu\beta=-\epsilon\varepsilon_\mu{}^\nu\partial_\nu\alpha$ where
$\varepsilon_\mu{}^\nu=\eta_{\mu\gamma}\varepsilon^{\gamma\nu}$,
$\varepsilon^{\mu\nu}=\left(\begin{array}{ll} 0& 1\\ -1&0\end{array}\right)$.
These geometrically defined functions can be used for construction
of a convenient pair of real null coordinates in the hyperbolic case or complex conjugated to each other coordinates in the elliptic case:
$$\begin{array}{l} \xi=\beta+j\alpha\\[1ex]
\eta=\beta-j\alpha\end{array}\qquad\qquad
j=\left\{\begin{array}{llll}
1,&\epsilon=1&-&\hbox{the hyperbolic case},\\[1ex]
i,&\epsilon=-1&- &\hbox{the elliptic case.}
\end{array}\right.$$

\subsection{Matter fields and their potentials}
The space-time symmetry described above implies that the following
gauge conditions can be imposed on the potentials $A_i$,  $B_i$
and $B_{ik}$:
$$A_\mu=0,\quad B_\mu=0,\quad B_{\mu\nu}=0,\quad B_{\mu a}=0$$
In this case, the fields $F_{ik}$ and $H_{ijk}$ are represented by
their non-vanishing components $A_a$, $B_{ab}$ depending on the
coordinates $x^\mu$ only. The antisymmetric matrix $B_{ab}$ can be
expressed in the form $B_{ab}= \Theta\,\epsilon_{ab}$, where
$\epsilon_{ab}$ is the antisymmetric Levi - Civita symbol. The
components $B_a$ and the scalar function $\Theta$ can be used also
instead of the components $A_a$ and the axion field $a$.
Thus, any of the field configuration under our consideration is
described by a set of dynamical variables
\begin{equation}\label{Variables}
g_{ab},\hskip1ex A_a\,(\hbox{or}\,B_a),\hskip1ex a\,(\hbox{or}\,\Theta),\hskip1ex\phi
\end{equation}
representing the components of metric,
gauge, axion and dilaton fields
respectively.

\section{Symmetry reduced field equations}

The dynamical part of the field equations of the string gravity
with the action (\ref{Action}), similarly to vacuum Einstein
equations ($2\times 2$-matrices) and electrovacuum Einstein -
Maxwell field equations ($3\times 3$-matrices) in General
Relativity, can be written in some complex $4\times 4$-matrix
form:
\begin{equation}\label{NullCurv}{\bf U}_\eta+{\bf
V}_\xi+\displaystyle\frac 1{2 i j\alpha} [{\bf U},{\bf V}] =
0,\qquad {\bf U}_\eta-{\bf V}_\xi=0,
\end{equation}
where $\alpha\equiv (\xi-\eta)/2 j$, and the complex $4\times 4$
matrices ${\bf U}$ and ${\bf V}$ should possess a universal (viz.
solution independent) canonical Jordan forms ${\bf U}_0$ and ${\bf
V}_0$:
\begin{equation}\label{Jordan}
{\bf U}_0=\hbox{diag}\,\{i,\,i,\,0,\,0\},\qquad {\bf
V}_0=\hbox{diag}\,\{i,\,i,\,0,\,0\}
\end{equation}
Besides that, the matrices ${\bf U}$ and ${\bf V}$ should satisfy
additional constraints. To describe these, we define at first a
Hermitian matrix ${\bf G}$ associated with every pair of matrices
${\bf U}$ and ${\bf V}$:
\begin{equation}\label{Gmatrix}
2 d{\bf G}={\bf \Omega}d {\cal U}-d{\cal
U}\,{}^\dagger {\bf \Omega},\qquad d {\cal U}\equiv {\bf
U}d\xi+{\bf V} d\eta,
\end{equation}
where ${}^\dagger$ stands for a Hermitian
conjugation of matrices (and matrix-valued 1-forms) and ${\bf
\Omega}$ is a constant matrix defined just below. Then the
corresponding algebraic constraints on ${\bf U}$ and ${\bf V}$ can
be expressed by the relations:
\begin{equation}\label{Constraints}
\begin{array}{lcl}
{\bf G}{\bf U}=i j\alpha{\bf \Omega}{\bf U}, && {\bf \Omega}{\bf U}+{\bf U}^T{\bf \Omega}=i {\bf \Omega},\\[1ex]
{\bf G}{\bf V}=-i j\alpha{\bf \Omega}{\bf V}, && {\bf \Omega}{\bf V}+{\bf V}^T{\bf \Omega}=i {\bf \Omega},
\end{array}\qquad{\bf \Omega}=\left(\begin{array}{cc} 0&I\\ -I& 0\end{array}\right),
\end{equation}
where ${}^T$ means a matrix transposition and $I$ denotes a
$2\times 2$ unit matrix. After an appropriate identification of the components of ${\bf U}$, ${\bf V}$ and ${\bf G}$ a direct calculations show that (\ref{NullCurv}) - (\ref{Constraints}) are equivalent to the reduced field equations for (\ref{Action}).

\section{Generalized (matrix) Ernst equation}

Similarly to the case of vacuum Einstein equations in General
Relativity, the equations (\ref{NullCurv}) - (\ref{Constraints})
can be reduced to one complex matrix equation for a symmetric
complex $2 \times 2$-matrix function, generalizing the well known
vacuum Ernst equation for a complex scalar Ernst potential and coinciding with 
such equation found in \cite{Kumar-Ray:1995}. For
this, using the upper right $2\times 2$-blocks ${\bf
U}_{(1)}{}^{(2)}$ and ${\bf V}_{(1)}{}^{(2)}$ of the $4\times
4$-matrices ${\bf U}$ and ${\bf V}$ we can define a generalized
matrix Ernst potential by the differential relation $d{\cal E}=-
{\bf U}_{(1)}{}^{(2)} d\xi-{\bf V}_{(1)}{}^{(2)} d\eta$. In the
most general form which includes both the hyperbolic and the
elliptic cases this matrix Ernst equation takes the
form
$$\left\{\begin{array}{l}
\eta^{\mu\nu}\left(\partial_\mu+\displaystyle\frac{\partial_\mu\alpha}\alpha
\right) \partial_\nu{\cal E}-\eta^{\mu\nu}\partial_\mu{\cal E}\cdot
(\hbox{Re}\,{\cal E})^{-1}\cdot\partial_\nu{\cal E}=0,\\[1ex]
\eta^{\mu\nu}\partial_\mu\partial_\nu \alpha=0 \end{array}\right.
$$
In the geometrically defined coordinates $(\xi,\eta)$ introduced in the previous section,
this equation takes a more simple form ($\alpha\equiv (\xi-\eta)/2 j$):
$${\cal E}_{\xi\eta}-\displaystyle\frac 1{4 j\alpha}({\cal E}_\xi-{\cal E}_\eta)-
\displaystyle\frac 12({\cal E}_\xi\cdot(\hbox{Re}\,{\cal E})^{-1}\cdot{\cal
E}_\eta + {\cal E}_\eta\cdot(\hbox{Re}\,{\cal E})^{-1}\cdot{\cal E}_\xi)=0
$$

\section{Equivalent spectral problem}
Generalizing the similar constructions for vacuum Einstein equations and electrovacuum Einstein - Maxwell field equations \cite{Alekseev:1985}-\cite{Alekseev:2001}, we consider the following spectral matrix problem for the four $4\times 4$-matrix functions ${\bf \Psi}(\xi,\eta,w)$, ${\bf U}(\xi,\eta)$, ${\bf V}(\xi,\eta)$ and ${\bf W}(\xi,\eta,w)$ ($w$ is a complex ``spectral'' parameter), which should satisfy the following three groups of conditions.
\begin{itemize}
\item{The overdetermined linear system for ${\bf \Psi}$ whose matrix coefficients
${\bf U}$, ${\bf V}$ should possess a universal (i.e. solution independent)  canonical forms:
$$
\left\{\begin{array}{l}
2 i(w-\xi)\partial_\xi{\bf \Psi}={\bf U}{\bf \Psi},\\
2 i(w-\eta)\partial_\eta{\bf \Psi}={\bf V}{\bf \Psi}
\end{array}\hskip1ex\right\Vert\hskip1ex
\left.\begin{array}{l}
{\bf U}={\cal F}_+ {\bf U}_0{\cal F}_+^{-1},\\
{\bf V}={\cal F}_- {\bf V}_0{\cal F}_-^{-1}
\end{array}\hskip1ex\right\Vert\hskip1ex
\begin{array}{l}
{\bf U}_0=\hbox{diag}\,\{i,\,i,\,0,\,0\},\\
{\bf V}_0=\hbox{diag}\,\{i,\,i,\,0,\,0\}.
\end{array}
$$
with some transformation matrices ${\cal F}_\pm$ depending on the field 
variables.}

\item{
This system should admit a Hermitian matrix
integral ${\bf K}(w)$:
$$
{\bf \Psi}^\dagger {\bf W}{\bf \Psi}={\bf K}(w), \qquad {\bf
K}^\dagger(w)={\bf K}(w) \qquad \displaystyle\frac {\partial {\bf
W}}{\partial w}=i{\bf \Omega}
$$
where ${\bf K}(w)$ is coordinate independent and ${\bf W}$ depends on the field variables.}
\item{
This system should admit also an antisymmetric
matrix integral ${\bf L}(w)$
$$\Sigma\cdot {\bf \Psi}^T {\bf \Omega}{\bf \Psi}={\bf L}(w),\qquad  {\bf L}^T(w)=-{\bf L}(w),\qquad  
\Sigma^2=(w-\xi)(w-\eta).
$$
where $\Sigma$ is an auxiliary scalar function.}
\end{itemize}
It can be shown directly (see, for the method details \cite{Alekseev:1988}), that solution of this spectral problem is equivalent to solution of (\ref{NullCurv}) - (\ref{Constraints}), and therefore, to solution of the symmetry reduced field equations for the string gravity model (\ref{Action}).

\section{Calculation of the field components}
In this concluding section we describe a relation 
between the reduced field equations and the spectral problem formulated above. For this we present here a general explicit expression of the components of the matrix ${\bf W}$ in terms of the field components: ${\bf W}\equiv i(w-\beta){\bf \Omega}+{\bf G}$ with

$${\bf G}=({\cal F}^{-1})^\dagger\left(\begin{array}{cccc}
-\epsilon_0 H\Omega^2-\epsilon_0 \displaystyle\frac{\epsilon\alpha^2} H&\epsilon_0 H\Omega&0&0\\
\epsilon_0 H\Omega&- \epsilon_0 H &0&0\\ 0&0& e^{-2\phi}\Theta^2+\epsilon\alpha^2e^{2\phi}&-e^{-2\phi}\Theta\\
0&0&- e^{-2\phi}\Theta&e^{-2\phi}
\end{array}\right){\cal F}^{-1}
$$
where we see two diagonal $2\times 2$ blocks which possess pure gravitational and axion-dilatonic nature respectively, while the ${\cal F}$-multipliers are of pure gauge nature:
$${\cal F}=\left(\begin{array}{cccc} 
1&0&\sqrt{2}A&0\\ 
0&0&1&0\\
0&1&\sqrt{2}\widetilde{A}&0\\ \sqrt{2}\widetilde{A}&
-\sqrt{2}A&0&1\end{array}\right).$$
We denote here $A_a=\{A,\tilde A\}$. It is useful to notice also that the spectral problem presented above possess an obvious gauge symmetry of the form
$$\begin{array}{lccl}
{\bf \Psi}\to{\bf A}\cdot{\bf \Psi}&&&
{\bf U}\to{\bf A}\cdot{\bf U}\cdot{\bf A}^{-1}\\
{\bf W}\to{{\bf A}^\dagger}^{-1}\cdot{\bf W}\cdot{\bf A}^{-1}&&&
{\bf V}\to{\bf A}\cdot{\bf V}\cdot{\bf A}^{-1}
\end{array}
$$
where the constant matrix ${\bf A}$ is real and satisfies the constraint ${\bf A}^T\cdot{\bf \Omega}\cdot {\bf A}={\bf \Omega}$.
However, this transformation possesses a nontrivial nature. In particular,  it can mix the mentioned above sectors and generate solutions with nonzero gauge fields $A_i$ from pure vacuum solutions.

\section{Acknowledgments}
This work is supported partly by the Russian Fund for Basic Researches (the grant 02-02-17372). The work of GAA is supported also by the Russian Fund for Basic Researches (the grant 02-01-00729) and by the programs ``Nonlinear dynamics'' of the Russian Academy of Sciences and ``Leading Scientific Schools'' of Russian Federation (the grant NSh-1697.2003.1).

\vfill
\end{document}